\documentclass[aps,prb,twocolumn,superscriptaddress,amsfonts,amssymb,showpacs]{revtex4}

\usepackage{graphicx}
\usepackage{dcolumn}
\usepackage{amssymb}
\usepackage{wasysym}

\begin{document}


\title{Transport and thermodynamic properties under anharmonic motion in type-I Ba$_{8}$Ga$_{16}$Sn$_{30}$ clathrate}


\author{Xiang Zheng}
\affiliation{Department of Physics and Astronomy, Texas A\&M University, College
Station, TX 77843, USA}
\affiliation{Materials Science and Engineering Program,  Texas A\&M University, College Station, TX 77843, USA}
\author{Sergio Y. Rodriguez}
\affiliation{Department of Physics and Astronomy, Texas A\&M University, College Station, TX 77843, USA}
\author{Laziz Saribaev}
\affiliation{Department of Physics and Astronomy, Texas A\&M University, College Station, TX 77843, USA}
\author{Joseph H. Ross, Jr.}
\affiliation{Department of Physics and Astronomy, Texas A\&M University, College Station, TX 77843, USA}
\affiliation{Materials Science and Engineering Program,  Texas A\&M University, College Station, TX 77843, USA}


\date{June 13, 2012}

\begin{abstract}
Anharmonic guest atom oscillation has direct connection to the thermal transport and thermoelectric behavior of type-I Ba$_{8}$Ga$_{16}$Sn$_{30}$ clathrates. This behavior can be observed through several physical properties, with for example the heat capacity providing a measure of the overall excitation level structure. Localized anharmonic excitations also influence the low-temperature resistivity, as we show in this paper. By combining heat capacity, transport measurements and  our previous NMR relaxation results, we address the distribution of local oscillators in this material, as well as the shape of the confining potential and the excitation energies for Ba(2) ions in the cages. We also compare to the soft-potential model and other models used for similar systems. The results show good agreement between the previously deduced anharmonic rattler potential and experimental data.
\end{abstract}

\maketitle

\section {INTRODUCTION}

Group IV clathrates are well known cage-structure materials with a single guest atom able to occupy each cage. Their outstanding thermoelectric performance and other potential useful properties have made them interesting for more than a decade \cite{NolasAPL1998, SnyderNature2008, NolasPRB2002, Dalton2010, AvilaPRB2006,SalesPRB2001}. Because of the loosely-held guest atoms, localized oscillators might be expected to represent a good model for their behavior, and this has been the focus of considerable research activity \cite{TakasuPRB2006,WeipingPRB2005}. In addition, a number of recent studies have shown that anharmonic phonon behavior may be a key element more generally for other types of thermoelectric materials \cite{DelaireNature2011, ZhangPRL2011}. In recent work, we have successfully analyzed the anharmonic motion for type-I Ba$_{8}$Ga$_{16}$Sn$_{30}$ using NMR relaxation measurements with a double well potential model \cite{XiangPRB2011}, while other methods, such as optical conductivity and first principles calculations, have also been utilized by other groups for similar systems \cite{MoriPRL2011, ZerecPRL2004, GeorgPRB2005, BiswasPRB2007}. 

Type-I Ba$_{8}$Ga$_{16}$Sn$_{30}$ clathrate is well-known for its ultra-low lattice thermal conductivity and glasslike thermal behavior, which are likely caused by the anharmonic rattling of the guest atoms inside the larger cages \cite{AvilaAPL2008, SuekuniPRB2010}. The well-known type-I clathrate structure features two structral cages, which for Ba$_{8}$Ga$_{16}$Sn$_{30}$ are each occupied by a Ba ion \cite{SalesPRB2001, BentienPRB2005}. The smaller cage is dodecahedral, and is occupied by the site designated Ba(1). Ba(2) occupies the larger cage, offering this ion considerably more space for vibrational motion. The cages themselves are formed by a connected network of four-bonded Ga-Sn sites.

The resistivity is very sensitive to electron-phonon coupling, and for the specific case of quasi-localized vibrational excitations, the resistivity can be a very useful analytic probe \cite{MahanPRB1993, CooperPRB1974}. Correspondingly, the heat capacity provides a measure to assess the density of the local oscillators, and the spacing of their energy levels \cite{MandrusPRB2001, HermannPRL2003, AvilaAPL2008}. The heat capacity has been examined in a number of recent studies of clathrate materials \cite{QiuPRB2004, UmeoJPSJ2005, SuekuniPRB2008, LortzPRB2008}. In this article, we investigate the influence on the resistivity as well as the heat capacity of the anharmonic rattling in type-I Ba$_{8}$Ga$_{16}$Sn$_{30}$ clathrate. Our previous NMR results will be used here to model the anharmonic potential \cite{XiangPRB2011}.

The previous NMR relaxation data, replotted in Fig. 1, were analyzed using a two-phonon Raman process according to a recent theory involving a localized one dimensional anharmonic oscillator potential \cite{DahmPRL2007}, shown also in the same figure. The results indicated that a relaxation mechanism due to anharmonic atomic motion is the leading contribution. The 1-D potential was thereby solved to give the calculated expression, $V(x)=-18.74 x^2+1.11\times10^{23} x^4$, where $V(x)$ is in J with $x$ given in m. The phonon frequency-temperature relationship and the energy levels of this double well potential can also offer corresponding ways to analyze the transport and heat capacity behavior \cite{XiangPRB2011}. In this paper we report measurements of this type and analyze for consistent behavior according to this model for the local anharmonic oscillators.
 
\section{Sample Preparation and Experimental Methods}

The type-I Ba$_{8}$Ga$_{16}$Sn$_{30}$ samples used here are the same as we used previously \cite{SergioMRS2010, XiangPRB2011} for structure configuration analysis (sample I-B from ref. 28), as well as for NMR relaxation studies \cite{XiangPRB2011}. Samples were prepared using the self-flux method, following a procedure reported previously \cite{SuekuniPRB2008, XiangPRB2011}. Pure elements were mixed based on the intended composition followed by an arc melting in argon. Annealing in an evacuated quartz tube at 900 $^{\rm o}$C for 50 hours was followed by a controlled slow cooling to 500 $^{\rm o}$C in 80 hours \cite{SuekuniPRB2008,SergioMRS2010}. Powder x-ray diffraction (XRD) was carried out using a Bruker D8 Advance diffractometer. Rietveld refinement of the results confirmed the structure to be type-I. No type-VIII reflections were detected and 1\% (per mol of corresponding framework atoms) Ba(Ga,Sn)$_4$ and Sn minority phases were also obtained \cite{SergioMRS2010}. Wavelength dispersion spectroscopy (WDS) measurements were carried out using a Cameca SX50 spectrometer, along with BaSO$_4$, GaP, and SnO$_2$ standards. The results correspond to a composition Ba$_{7.8(1)}$Ga$_{16.2(1)}$Sn$_{29.9(1)}$, where errors correspond to the statistical errors. Taken together these results are consistent and point to an electron-deficient p-type composition, relative to an electron-balanced Zintl phase. All resistivity and heat capacity measurements were carried out using a Quantum Design Physical Property Measurement System (PPMS).

\begin{figure}
\includegraphics[width=\columnwidth]{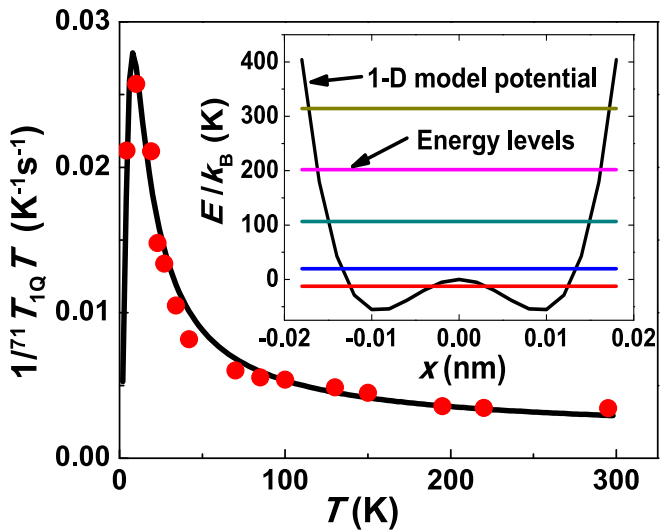}
\caption{\label{fig:fig1}  (Color online) $^{71}$Ga NMR quadrupole contribution to inverse $T_1T$ product  (circles) from ref. 9 for  type-I Ba$_{8}$Ga$_{16}$Sn$_{30}$, and a fit to anharmonic oscillator behavior (solid curve). Inset: Corresponding 1-D double well potential and its first few energy levels \cite{XiangPRB2011}.}
\end{figure}

\section{Resistivity}

The resistivity data are plotted in Fig. 2. Note that there is a superconducting jump at about 4 K, which can be associated with the Sn minor phase observed in the XRD result. In this work, only the non-superconducting part will be shown and analyzed.

For a system with sufficient carrier density to be metallic, and assuming the majority of the electrical resistivity is caused by the ordinary electron-phonon interaction and follows the standard Bloch-Gr{\"u}neisen law \cite{Bloch-Gruneisen-Ziman},
\begin{eqnarray}
{\rho}_B(T)={\rho}_0+A\left(\frac{T}{{\Theta}_D}\right)^5{\int}_0^{{\Theta}_D/T}\frac{x^5dx}{(e^x-1)(1-e^{-x})},
\end{eqnarray}
where ${\Theta}_D$ is the Debye temperature, ${\rho}_0$ is the residual resistivity and $A$ is a constant. This is not sufficient for a system with localized harmonic and anharmonic oscillators. According to Cooper's theory \cite{CooperPRB1974}, the Einstein contribution is proportional to $C_ET/\Theta_E^2$ as
\begin{eqnarray}
{\rho}_E(T)=\frac{{\alpha}C_ET}{\Theta_E^2}=\left(\frac{\kappa}{T}\right)\frac{e^{{\Theta}_E/T}}{(e^{{\Theta}_E/T}-1)^2},
\end{eqnarray}
where $\alpha$ and $\kappa$ are constants, $C_E$ is the Einstein contribution to the specific heat and ${\Theta}_E$ is the Einstein temperature. Since Ba atoms exist in two different types of cages, we can consider two different oscillator behaviors of this type. These local modes are resonances within the phonon bands, however the localized model works relatively well, implying a weak coupling to other lattice modes.

We started with a fit including one Bloch-Gr{\"u}neisen term and two Einstein terms with the results shown in Fig. 2. Here, we define ${\Theta}_{E1}$ and ${\Theta}_{E2}$ as the Einstein temperatures for Ba(1) and Ba(2) atoms.  The fitted parameters from standard deviation calculations are  ${\Theta}_D$ = 230 K, ${\Theta}_{E1}$ = 56 K, ${\Theta}_{E2}$ = 49 K and ${\rho}_0$ = 243 ${\mu}{\Omega}$ cm. This gives noticeable improvement over the fit with a single Bloch-Gr{\"u}neisen contribution (not shown). The Bloch-Gr{\"u}neisen term is appropriate for metallic systems, and our previous NMR results \cite{XiangPRB2011} showed that a Korringa-like behavior (constant magnetic NMR shift and $T_{1M}T$ nearly $T$-independent) is followed in the material, which is a sign of metallic behavior. 

The overall fit vs. temperature matches particularly well at high temperature, but the inset of Fig. 2 shows a mis-match at the low temperature end. Previous studies have shown a $T^2$ resistivity behavior in low temperature caused by anharmonic phonons \cite{DahmPRL2007, MahanPRB1993}, which is close to what is observed here. For example, fitting the data up to 12 K to a function of the form $T^\alpha$ gives $\alpha=2.2$. An alternative explanation for the deviation from $T^5$ resistivity behavior at low temperatures might be semiconducting behavior as expected in low-carrier density systems. For example in non-polar semiconductors \cite{BulusuSM2008} acoustic phonon scattering can introduce a term proportional to $T^{1.5}$. On the other hand, as described above, the  NMR shifts and magnetic $T_1$ term provide additional local measures to confirm that the carrier behavior should be regarded as metallic \cite{XiangPRB2011}. Combined with the strong evidence for anharmonic rattling observed in the NMR relaxation results, which becomes evident in the same range of temperatures, it seems reasonable to model the observed resistivity behavior according to the  anharmonic behavior of the guest Ba(2) atom.

\begin{figure}
\includegraphics[width=\columnwidth]{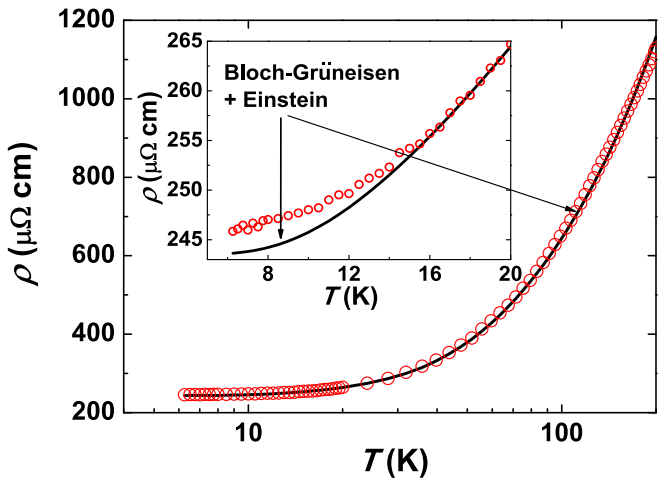}
\caption{\label{fig:fig2} (Color online) Resistivity measurements (open circles) and fitting (solid curve) from Bloch-Gr{\"u}neisen function and Einstein model with ${\Theta}_D=230$ K and  ${\Theta}_{E1}=56$ K, ${\Theta}_{E2}=49$ K. Inset: Expanded view at the low temperature end of the data and fitting, where there is a clear mis-match between the model and data. }
\end{figure}

The electrical resistivity due to localized anharmonic phonons has been addressed in recent theoretical work \cite{DahmPRL2007} and can be calculated from the electron lifetime ($\tau$), which describes the electron scattering from phonons \cite{Bloch-Gruneisen-Ziman,Kittel}. The corresponding resistivity is given by,
\begin{eqnarray}
{\rho}_A(T)=\frac{m^*}{n_0e^2{\tau}(T)}.
\end{eqnarray} 
The temperature-dependent electron lifetime (${\tau}(T)$) can be obtained by averaging the energy-dependent lifetime,
\begin{eqnarray}
{\tau}(T)={\int}_{\infty}^{\infty}dE{\tau}(E)\left(-\frac{df(E)}{dE}\right),
\end{eqnarray}
where $f(E)=\frac{1}{{\rm exp}\{E/k_BT\}+1}$ is the Fermi function. Furthermore, ${\tau}(E)$ can be obtained from the imaginary part of the retarded self-energy \cite{MahanPRB1993,DahmPRL2007},
\begin{eqnarray}
\tau^{-1}(E)={\pi}g^2N(0){\int}_0^{\infty}d{\Omega}A(\Omega)[2n(\Omega)+f({\hbar}\Omega+E)\nonumber\\+f({\hbar}\Omega-E)]
\end{eqnarray}
where $A(\Omega)=-\frac{1}{\pi}ImD(\Omega)=\frac{1}{\pi}\frac{4{\omega}_0{\Gamma}_0\Omega}{({\Omega}^2-{\omega}_r^2)^2+4{\Gamma}_0^2{\Omega}^2}$
is assumed to be the phonon spectral function. The effective localized phonon frequency ${\omega}_0$, phonon damping rate ${\Gamma}_0$ and renormalized phonon frequency ${\omega}_r$ are all defined in this way as reported before \cite{DahmPRL2007,XiangPRB2011}. The previously reported NMR results yielded a large damping coefficient, ${\Gamma}_0$ = 12 K, which will tend to enhance this mechanism at low temperatures. Note that a damped 1D anharmonic model was considered in an analysis of the optical conductivity \cite{MoriPRL2011}, yielding a damping coefficient $\Gamma\approx0.5$ THz = 24 K at low temperatures, not far from the value we reported. For $T << \hbar \omega_r/k_B$ , the calculated  resistivity will follow a ${\rho}$ ${\sim}$ $T^2$ relationship as described above \cite{DahmPRL2007}. Thus we examine a combination of the Bloch-Gr{\"u}neisen function, Einstein model and  anharmonic model with respect to the resistivity in the low temperature region. 
 
\begin{figure}
\includegraphics[width=\columnwidth]{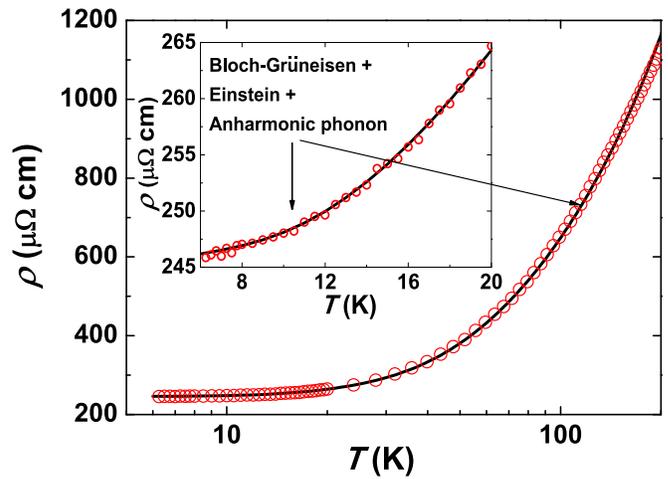}
\caption{\label{fig:fig3} (Color online) Resistivity data (open circles) and fitting (solid curve) with rattler contribution added to the previous model, ${\Theta}_D=230$ K, ${\Theta}_{E1}=66$ K and ${\Theta}_{E2}=54$ K. Inset: Expanded view of the low temperature region. The mis-match between the data and the model has been reduced significantly compared to that of Fig. 2}
\end{figure}

In fitting the resistivity, we use the model for ${\Theta}_D$, ${\Theta}_{E1}$, ${\Theta}_{E2}$, ${\rho}_0$ as above with the addition of an anharmonic contribution with the same damping rate ${\Gamma}_0$ and temperature dependent phonon frequencies (${\omega}_0$, ${\omega}_r$) as our previous NMR results \cite{XiangPRB2011}. Fig. 3 shows the result from this combined model with ${\Theta}_D=230$ K, ${\Theta}_{E1} = 66$ K, ${\Theta}_{E2}$ = 54K, ${\rho}_0$ = 245 ${\mu}{\Omega}$ cm. A single additional parameter represents the overall strength of the anharmonic contribution. The high-temperature agreement remains as good as that in Fig. 2, but the inset of Fig. 3 shows a much improved fit in the low temperature region. Note that the anharmonic portion only exhibits a strong contribution at low temperatures. We emphasize that this combined model starts directly from specific physical mechanisms in this system, so the results should be consistent with heat capacity as we examine below.

\section{Heat Capacity}

Heat capacity data from 2 K to 200 K are shown in Fig. 4 with a fit including several mechanisms as described below. In fitting the data, the leading contribution was taken as a Debye model for the framework atoms,
\begin{eqnarray}
C_D=9N_DR(\frac{T}{{\Theta}_D})^3{\int}_0^{{\Theta}_D/T}\frac{x^4e^xdx}{(e^x-1)^2},
\end{eqnarray}
where ${\Theta}_D$ is the Debye temperature, and $N_D$ is fixed at 46, the number of framework atoms per cell.

In a similar way as for the resistivity, the six Ba(2) atoms are considered to be rattlers with both anharmonic and harmonic motions corresponding to the different directions. We assume the anharmonicity to be active in one direction, so the simulation will start with six 1D anharmonic oscillators and six 2D Einstein oscillators for these atoms. We use ${\Theta}_{E2}$ as their Einstein temperature, $N_{E2}$ as the 2D Einstein oscillator number and $N_{Anh}$ as the 1D anharmonic oscillator number. Two Ba(1) atoms, inside the smaller cages, are treated by a 3D Einstein model with parameters ${\Theta}_{E1}$ and $N_{E1}$. These follow the standard behavior
\begin{eqnarray}
C_E=3N_ER\left(\frac{{\Theta}_E}{T}\right)^2\frac{e^{{\Theta}_E/T}}{(e^{{\Theta}_E/T}-1)^2}.
\end{eqnarray}
For the anharmonic contribution, we used
\begin{eqnarray}
U=\frac{\sum\limits_{n=0}^{\infty}E_n{\rm exp}\{-E_n/k_BT\}}{\sum\limits_{n=0}^{\infty}{\rm exp}\{-E_n/k_BT\}}, \ C_A=\frac{dU}{dT},
\end{eqnarray}
where the energy levels $E_n$, shown in the inset of Fig. 1,  are those corresponding to the anharmonic potential. We generated these by solving the Schr{\"o}dinger equation numerically with the 1-D double-well parameters from our previous NMR results. The lowest 13 levels were used, after we verified that higher levels add a sufficiently small contribution to the sum to be ignored. 

\begin{figure}
\includegraphics[width=\columnwidth]{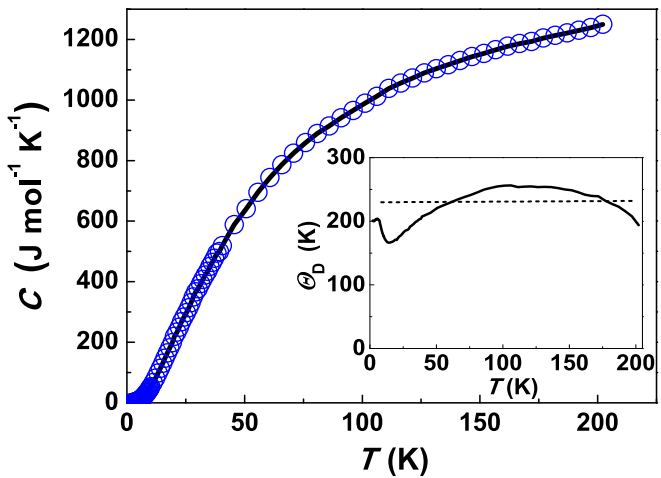}
\caption{\label{fig:fig4} (Color online) Heat capacity measurement (open circles) and fitting (solid curve) to the model discribed in text. Inset: Temperature dependent $\Theta_D(T)$ (solid curve). For comparison, the Debye temperature from the resistivity fit is also shown here (dashed line).}
\end{figure}

The fitting result shown in Fig. 4 and 5 gives $N_{E1}$ = 2, ${\Theta}_{E1}$ = 70 K, $N_{E2}$= 6, ${\Theta}_{E2}$ = 55 K, $N_{Anh}$ = 5.4. Also, we used a temperature dependent Debye temperature $\Theta_D(T)$ for the fitting, resulting in a typical behavior as shown in the inset of Fig. 4, with values near 230 K. While carrier scattering depends upon the phonon mode so that $\Theta_D$ extracted from resistivity need not be equal to the heat capacity-related $\Theta_D$ \cite{Grimvall}, often these values are quite close, as observed here. See comparison in Table.~\ref{Comparison} of this and other parameters from the fitting. The electronic contribution, $\gamma$T, is fitted to $\gamma=1.85$ mJ/mol K$^2$ for each atom. Note that a reduced number of anharmonic oscillators is obtained, with about 10\% missing, relative to the expected 6 per cell. To account for the missing oscillator strength and the observed low temperature tail in $C$, we added a low energy Einstein term with $N_{E3}$ = 0.6, $\Theta_{E3}=14.2$ K.  The $C/T^3$ vs. $T$ plot in Fig. 5 shows the contribution of each term. The Einstein part is a superposition of the three Einstein terms. With the exception of the small $\Theta_{E3}$ term, the fitted results are in good agreement with those obtained from resistivity. Notice that, $\Theta_{E3}$ term only contributes significantly below 5 K, so it will not introduce noticeable influence to the resistivity fit. The broad peak in $C/T^3$ at low temperatures agrees well with the anharmonic parameters taken directly from the NMR fit, and the result serves to quantify the corresponding number of anharmonic oscillators. The model dividing the oscillator strength into localized and extended parts thus provides a consistent explanation for these results. The model and analysis work well for NMR, resistivity and heat capacity for this sample. Further investigations with samples of different compositions will offer additional understanding of the mechanism. 

\begin{table}
\caption{Comparison of fitted parameters from resistivity and heat capacity analysis.}
\begin{center}
\label{Comparison}
\begin{ruledtabular}
\begin{tabular}{  l   c   c   c   c }
 & $\Theta_{E1}$(K)  &  $\Theta_{E2}$(K)  &   $\Theta_{E3}$(K)  &  $\Theta_{D}$(K) \\
\hline
Resistivity  & 66 & 54 & -- & 230  \\
Heat capacity & 70 & 55 & 14.2 & [170-260] \\
\end{tabular}
\end{ruledtabular}
\end{center}
\end{table}

\begin{figure}
\includegraphics[width=\columnwidth]{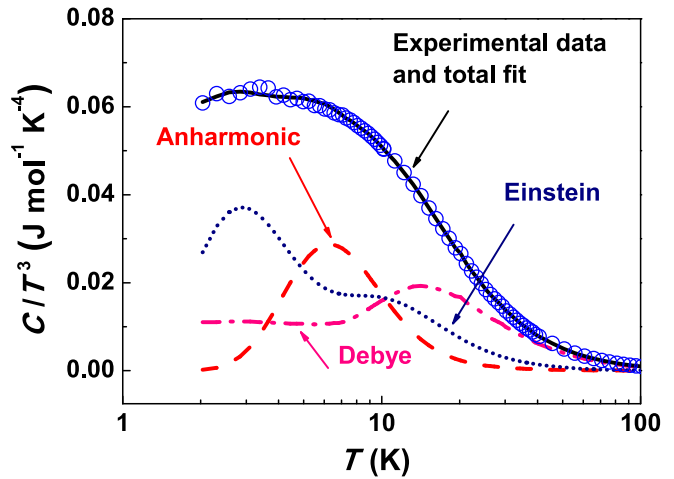}
\caption{\label{fig:fig5}  Measured $C/T^3$ vs. $T$ and the fit to model discribed in text (solid curve). Individual contributions as labeled: Debye (dash-dotted), Einstein (dotted) and anharmonic oscillator (dashed). The Einstein part is a superposition of several oscillators.}
\end{figure}

\section{Discussion}

Often for modeling of the heat capacity in clathrate systems a multi-Einstein model is used to describe the broad distribution representing the low temperature peak in $C/T^3$ vs. $T$. This works reasonably well for Ba$_{8}$Ga$_{16}$Ge$_{30}$ and Sr$_{8}$Ga$_{16}$Ge$_{30}$ among others \cite{AvilaPRB2006, SuekuniPRB2008}. However, Ba$_{8}$Ga$_{16}$Sn$_{30}$ exhibits a broad peak, and correspondingly the large cage-center position is marginally stable \cite{AvilaAPL2008} or perhaps unstable to off-center ion displacements \cite{MoriPRL2011}, which suggests an anharmonic rattling model as has been applied to other results. Our analysis shows that a specific local potential of this type can be connected to several experimental results in a consistent way, thus providing a good physical picture for the vibrational behavior. The large damping coefficient indicated by the resistivity as well as the NMR results implies that these modes are strongly coupled to other excitations, and thus cannot be regarded as completely independent oscillators. Recent research on phonon dispersion in clathrates including X$_{8}$Ga$_{8}$Ge$_{128}$, X$_{8}$Ga$_{16}$Si$_{128}$, Rb$_{2}$Sr$_6$Ga$_{14}$Ge$_{32}$ among others, have shown strong interactions between localized rattler modes and the framework atoms \cite{BiswasPRB2007, DongJAP2000, RenyPRB2002, HermannPRB2005, JohnsonPRB2010, ChistensenNatMater2008, ChristensenJACS2006, RoudebushInorgChem2012, LeeJPSJ2008}, which may offer an explanation for this phenomenon.

Our fitting works surprisingly well based on an initial assumption that the anharmonic motion is one dimensional, giving six 1D anharmonic oscillators. This differs from the expected two dimensional behavior, often as a four-well potential due to the configuration of the Ba(2) cages \cite{ZerecPRL2004}. However recent studies point to an off-center symmetry for Ba(2) oscillations \cite{AvilaAPL2008}, and our previous $ab$ $initio$ results \cite{SergioMRS2010} indicate a static off-center displacement of as much as 0.5 {\AA} for Ba on this site based on Ga-Sn alloy disorder. With sufficient cage distortion, rattling-type vibrations near the cage minimum could be constrained to be effectively one-dimensional, with a harmonic restoring potential in other directions. Our previous report on atom configuration and bond length calculations also pointed out possible structural distortions for this material \cite{SergioMRS2010}. An alternative view might be that among the two-dimensional anharmonic oscillators, approximately half of the rattlers are not activated, accounting for the corresponding missing spectral weight from the heat capacity fit. It might be that the presence of stronger defects, such as vacancies, leads to this situation in some of the cages. However we also note that a fit of our heat capacity data using a 2-D rotationally symmetric anharmonic potential as was also fitted to the NMR relaxation results \cite{XiangPRB2011}, did not work well and placed the heat capacity peak at temperatures too high for reasonable agreement. 

We should remark that both the WDS and XRD results show a small reduction of the Ba atom content relative to the stoichiometric composition. However this amount is much too small to account for any significant discrepancy in number of rattler atoms. According to WDS measurements at several places in the ingot \cite{SergioMRS2010} the Ba content is reduced by about 2\%. Given the measured small Ga excess, this sample would be expected to be p type \cite{SuekuniPRB2008}. Such a composition would also not be expected to exhibit a large number of spontaneous vacancies, as they are not needed to maintain the Zintl electron count.

\begin{figure}
\includegraphics[width=\columnwidth]{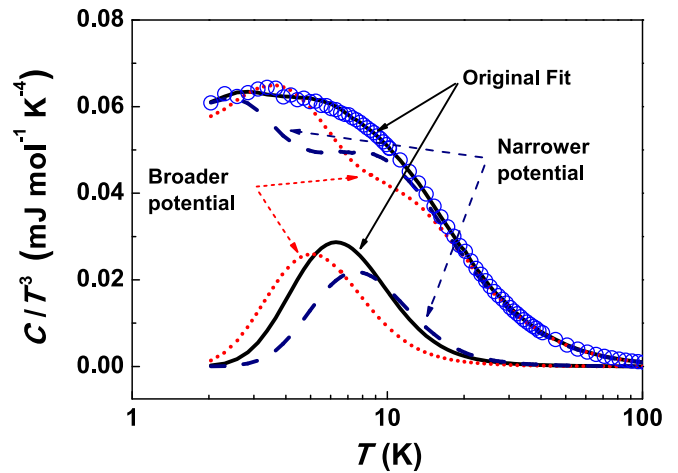}
\caption{(Color online) Comparison of the anharmonic contribution and total calculated heat capacity with different anharmonic potentials. Compared to the model based on previous NMR results (solid curves), the dotted curves represent a narrower anharmonic potential and the dashed curves represent a broader anharmonic potential, as described in the text.}
\label{shift}  
\end{figure}

Another well known approximation, the soft potential model (SPM), has been introduced into the analysis of heat capacity in many systems with anharmonic contributions \cite{RamosSPM,UmeoJPSJ2005,SuekuniPRB2008,LindqvistJCP1997}. This is based on a soft vibrational density of states and the tunneling of an assumed wide distribution of two level systems. A significant contribution from two level tunneling systems was deduced in Ba$_{8}$Ga$_{16}$Sn$_{30}$ and Sr$_{8}$Ga$_{16}$Ge$_{30}$ at low temperatures \cite{UmeoJPSJ2005, SuekuniPRB2008}. A very broad distribution of oscillation frequencies are considered in this model, which can simulate both the anharmonic contribution and harmonic contributions. Indeed, the small added Einstein term at low temperatures in the heat capacity fit may represent a distribution of tunnel sites of this type. It is not clear what may be the origin of these additional tunneling systems, however their number is relatively low. Thus while the SPM model alone does work reasonably well in analyzing the heat capacity, we believe that the results shown here point to a strongly damped anharmonic potential as a more physical model for this system.

The sensitivity of the overall fit to the fitting parameters is also important to discuss. For the heat capacity, the fitted oscillator numbers are important, because they not only determine the overall shape of the fit, but also offer a physical picture of the localized motions. For the anharmonic contribution, a variation in the anharmonicity affects the energy levels, changing the position as well as magnitude of the anharmonic contribution to the $C/T^3$ plot. In the original fit we used anharmonic well parameters taken directly from the reported NMR results, however Fig.~\ref{shift} shows results for which the anharmonic potential well width was scaled by $\pm$10\%, without changing its shape. This corresponds to a scaling of the energy levels by $\pm$20\%. For these curves, the Debye and Einstein temperatures were not changed, however the oscillator numbers were allowed to adjust, with a result that the numbers no longer match the composition, and the agreement with the measured curve is clearly made worse. This indicates the sensitivity of the fit to the anharmonic potential. It is possible to obtain an improved agreement with such a scaled potential by allowing the Einstein temperatures to change, however this occurs by shifting the lowest $\Theta_{E}$ values on top of the anharmonic peak, a result that is similar to the SPM model discussed above, in which a distribution of harmonic oscillators approximates the distribution of energy levels.

\section{Conclusions}

We have shown that our resistivity, heat capacity and NMR results are consistent with the anharmonic rattling model in this type-I Ba$_{8}$Ga$_{16}$Sn$_{30}$ clathrate. We utilized an $x^4$-type anharmonic potential, which provides good agreement between the NMR results, heat capacity, and transport measurements, with a single set of parameters. The damping parameter is large, indicating that these vibrations interact strongly with vibrational or electronic excitations in the framework. However the success of the 1-D model in this case implies that the expected 2-D motion of these rattlers is not activated, perhaps through cage distortion.

\begin{acknowledgments}
This work was supported by the Robert A. Welch Foundation, Grant
No. A-1526, by the National Science Foundation (DMR-0103455).
\end{acknowledgments}


\begin{thebibliography}{42}
\expandafter\ifx\csname natexlab\endcsname\relax\def\natexlab#1{#1}\fi
\expandafter\ifx\csname bibnamefont\endcsname\relax
  \def\bibnamefont#1{#1}\fi
\expandafter\ifx\csname bibfnamefont\endcsname\relax
  \def\bibfnamefont#1{#1}\fi
\expandafter\ifx\csname citenamefont\endcsname\relax
  \def\citenamefont#1{#1}\fi
\expandafter\ifx\csname url\endcsname\relax
  \def\url#1{\texttt{#1}}\fi
\expandafter\ifx\csname urlprefix\endcsname\relax\def\urlprefix{URL }\fi
\providecommand{\bibinfo}[2]{#2}
\providecommand{\eprint}[2][]{\url{#2}}

\bibitem[{\citenamefont{Nolas et~al.}(1998)\citenamefont{Nolas, Cohn, Slack,
  and Schujman}}]{NolasAPL1998}
\bibinfo{author}{\bibfnamefont{G.~S.} \bibnamefont{Nolas}},
  \bibinfo{author}{\bibfnamefont{J.~L.} \bibnamefont{Cohn}},
  \bibinfo{author}{\bibfnamefont{G.~A.} \bibnamefont{Slack}}, \bibnamefont{and}
  \bibinfo{author}{\bibfnamefont{S.~B.} \bibnamefont{Schujman}},
  \bibinfo{journal}{Appl. Phys. Lett.} \textbf{\bibinfo{volume}{73}},
  \bibinfo{pages}{178} (\bibinfo{year}{1998}).

\bibitem[{\citenamefont{Snyder and Toberer}(2008)}]{SnyderNature2008}
\bibinfo{author}{\bibfnamefont{G.~J.} \bibnamefont{Snyder}} \bibnamefont{and}
  \bibinfo{author}{\bibfnamefont{E.~S.} \bibnamefont{Toberer}},
  \bibinfo{journal}{Nat. Mater.} \textbf{\bibinfo{volume}{7}},
  \bibinfo{pages}{105} (\bibinfo{year}{2008}).

\bibitem[{\citenamefont{Nolas et~al.}(2002)\citenamefont{Nolas, Cohn, Dyck,
  Uher, and Yang}}]{NolasPRB2002}
\bibinfo{author}{\bibfnamefont{G.~S.} \bibnamefont{Nolas}},
  \bibinfo{author}{\bibfnamefont{J.~L.} \bibnamefont{Cohn}},
  \bibinfo{author}{\bibfnamefont{J.~S.} \bibnamefont{Dyck}},
  \bibinfo{author}{\bibfnamefont{C.}~\bibnamefont{Uher}}, \bibnamefont{and}
  \bibinfo{author}{\bibfnamefont{J.}~\bibnamefont{Yang}},
  \bibinfo{journal}{Phys. Rev. B} \textbf{\bibinfo{volume}{65}},
  \bibinfo{pages}{165201} (\bibinfo{year}{2002}).

\bibitem[{\citenamefont{Christensen et~al.}(2010)\citenamefont{Christensen,
  Johnsen, and Iversen}}]{Dalton2010}
\bibinfo{author}{\bibfnamefont{M.}~\bibnamefont{Christensen}},
  \bibinfo{author}{\bibfnamefont{S.}~\bibnamefont{Johnsen}}, \bibnamefont{and}
  \bibinfo{author}{\bibfnamefont{B.~B.} \bibnamefont{Iversen}},
  \bibinfo{journal}{Dalton Trans.} \textbf{\bibinfo{volume}{39}},
  \bibinfo{pages}{978} (\bibinfo{year}{2010}).

\bibitem[{\citenamefont{Avila et~al.}(2006)\citenamefont{Avila, Suekuni, Umeo,
  Fukuoka, Yamanaka, and Takabatake}}]{AvilaPRB2006}
\bibinfo{author}{\bibfnamefont{M.~A.} \bibnamefont{Avila}},
  \bibinfo{author}{\bibfnamefont{K.}~\bibnamefont{Suekuni}},
  \bibinfo{author}{\bibfnamefont{K.}~\bibnamefont{Umeo}},
  \bibinfo{author}{\bibfnamefont{H.}~\bibnamefont{Fukuoka}},
  \bibinfo{author}{\bibfnamefont{S.}~\bibnamefont{Yamanaka}}, \bibnamefont{and}
  \bibinfo{author}{\bibfnamefont{T.}~\bibnamefont{Takabatake}},
  \bibinfo{journal}{Phys. Rev. B} \textbf{\bibinfo{volume}{74}},
  \bibinfo{pages}{125109} (\bibinfo{year}{2006}).

\bibitem[{\citenamefont{Sales et~al.}(2001)\citenamefont{Sales, Chakoumakos,
  Jin, Thompson, and Mandrus}}]{SalesPRB2001}
\bibinfo{author}{\bibfnamefont{B.~C.} \bibnamefont{Sales}},
  \bibinfo{author}{\bibfnamefont{B.~C.} \bibnamefont{Chakoumakos}},
  \bibinfo{author}{\bibfnamefont{R.}~\bibnamefont{Jin}},
  \bibinfo{author}{\bibfnamefont{J.~R.} \bibnamefont{Thompson}},
  \bibnamefont{and} \bibinfo{author}{\bibfnamefont{D.}~\bibnamefont{Mandrus}},
  \bibinfo{journal}{Phys. Rev. B} \textbf{\bibinfo{volume}{63}},
  \bibinfo{pages}{245113} (\bibinfo{year}{2001}).

\bibitem[{\citenamefont{Takasu et~al.}(2006)\citenamefont{Takasu, Hasegawa,
  Ogita, Udagawa, Avila, Suekuni, Ishii, Suzuki, and
  Takabatake}}]{TakasuPRB2006}
\bibinfo{author}{\bibfnamefont{Y.}~\bibnamefont{Takasu}},
  \bibinfo{author}{\bibfnamefont{T.}~\bibnamefont{Hasegawa}},
  \bibinfo{author}{\bibfnamefont{N.}~\bibnamefont{Ogita}},
  \bibinfo{author}{\bibfnamefont{M.}~\bibnamefont{Udagawa}},
  \bibinfo{author}{\bibfnamefont{M.~A.} \bibnamefont{Avila}},
  \bibinfo{author}{\bibfnamefont{K.}~\bibnamefont{Suekuni}},
  \bibinfo{author}{\bibfnamefont{I.}~\bibnamefont{Ishii}},
  \bibinfo{author}{\bibfnamefont{T.}~\bibnamefont{Suzuki}}, \bibnamefont{and}
  \bibinfo{author}{\bibfnamefont{T.}~\bibnamefont{Takabatake}},
  \bibinfo{journal}{Phys. Rev. B} \textbf{\bibinfo{volume}{74}},
  \bibinfo{pages}{174303} (\bibinfo{year}{2006}).

\bibitem[{\citenamefont{Gou et~al.}(2005)\citenamefont{Gou, Li, Chi, Ross,
  Beekman, and Nolas}}]{WeipingPRB2005}
\bibinfo{author}{\bibfnamefont{W.}~\bibnamefont{Gou}},
  \bibinfo{author}{\bibfnamefont{Y.}~\bibnamefont{Li}},
  \bibinfo{author}{\bibfnamefont{J.}~\bibnamefont{Chi}},
  \bibinfo{author}{\bibfnamefont{J.~H.} \bibnamefont{Ross},
  \bibfnamefont{Jr.}},
  \bibinfo{author}{\bibfnamefont{M.}~\bibnamefont{Beekman}}, \bibnamefont{and}
  \bibinfo{author}{\bibfnamefont{G.~S.} \bibnamefont{Nolas}},
  \bibinfo{journal}{Phys. Rev. B} \textbf{\bibinfo{volume}{71}},
  \bibinfo{pages}{174307} (\bibinfo{year}{2005}).

\bibitem[{\citenamefont{Delaire et~al.}(2011)\citenamefont{Delaire, J.~Ma, May,
  McGuire, Du, Singh, Podlesnyak, Ehlers, Lumsden, and
  Sales}}]{DelaireNature2011}
\bibinfo{author}{\bibfnamefont{O.}~\bibnamefont{Delaire}},
  \bibinfo{author}{\bibfnamefont{K.~M.} \bibnamefont{J.~Ma}},
  \bibinfo{author}{\bibfnamefont{A.~F.} \bibnamefont{May}},
  \bibinfo{author}{\bibfnamefont{M.~A.} \bibnamefont{McGuire}},
  \bibinfo{author}{\bibfnamefont{M.-H.} \bibnamefont{Du}},
  \bibinfo{author}{\bibfnamefont{D.~J.} \bibnamefont{Singh}},
  \bibinfo{author}{\bibfnamefont{A.}~\bibnamefont{Podlesnyak}},
  \bibinfo{author}{\bibfnamefont{G.}~\bibnamefont{Ehlers}},
  \bibinfo{author}{\bibfnamefont{M.~D.} \bibnamefont{Lumsden}},
  \bibnamefont{and} \bibinfo{author}{\bibfnamefont{B.~C.} \bibnamefont{Sales}},
  \bibinfo{journal}{Nat. Mater.} \textbf{\bibinfo{volume}{10}},
  \bibinfo{pages}{614} (\bibinfo{year}{2011}).

\bibitem[{\citenamefont{Zhang et~al.}(2011)\citenamefont{Zhang, Ke, Kent, Yang,
  and Chen}}]{ZhangPRL2011}
\bibinfo{author}{\bibfnamefont{Y.}~\bibnamefont{Zhang}},
  \bibinfo{author}{\bibfnamefont{X.}~\bibnamefont{Ke}},
  \bibinfo{author}{\bibfnamefont{P.~R.~C.} \bibnamefont{Kent}},
  \bibinfo{author}{\bibfnamefont{J.}~\bibnamefont{Yang}}, \bibnamefont{and}
  \bibinfo{author}{\bibfnamefont{C.}~\bibnamefont{Chen}},
  \bibinfo{journal}{Phys. Rev. Lett.} \textbf{\bibinfo{volume}{107}},
  \bibinfo{pages}{175503} (\bibinfo{year}{2011}).

\bibitem[{\citenamefont{Zheng et~al.}(2011)\citenamefont{Zheng, Rodriguez, and
  Ross}}]{XiangPRB2011}
\bibinfo{author}{\bibfnamefont{X.}~\bibnamefont{Zheng}},
  \bibinfo{author}{\bibfnamefont{S.~Y.} \bibnamefont{Rodriguez}},
  \bibnamefont{and} \bibinfo{author}{\bibfnamefont{J.~H.} \bibnamefont{Ross},
  \bibfnamefont{Jr.}}, \bibinfo{journal}{Phys. Rev. B}
  \textbf{\bibinfo{volume}{84}}, \bibinfo{pages}{024303}
  (\bibinfo{year}{2011}).

\bibitem[{\citenamefont{Mori et~al.}(2011)\citenamefont{Mori, Iwamoto,
  Kushibiki, Honda, Matsumoto, Toyota, Avila, Suekuni, and
  Takabatake}}]{MoriPRL2011}
\bibinfo{author}{\bibfnamefont{T.}~\bibnamefont{Mori}},
  \bibinfo{author}{\bibfnamefont{K.}~\bibnamefont{Iwamoto}},
  \bibinfo{author}{\bibfnamefont{S.}~\bibnamefont{Kushibiki}},
  \bibinfo{author}{\bibfnamefont{H.}~\bibnamefont{Honda}},
  \bibinfo{author}{\bibfnamefont{H.}~\bibnamefont{Matsumoto}},
  \bibinfo{author}{\bibfnamefont{N.}~\bibnamefont{Toyota}},
  \bibinfo{author}{\bibfnamefont{M.~A.} \bibnamefont{Avila}},
  \bibinfo{author}{\bibfnamefont{K.}~\bibnamefont{Suekuni}}, \bibnamefont{and}
  \bibinfo{author}{\bibfnamefont{T.}~\bibnamefont{Takabatake}},
  \bibinfo{journal}{Phys. Rev. Lett.} \textbf{\bibinfo{volume}{106}},
  \bibinfo{pages}{015501} (\bibinfo{year}{2011}).

\bibitem[{\citenamefont{Zerec et~al.}(2004)\citenamefont{Zerec, Keppens,
  McGuire, Mandrus, Sales, and Thalmeier}}]{ZerecPRL2004}
\bibinfo{author}{\bibfnamefont{I.}~\bibnamefont{Zerec}},
  \bibinfo{author}{\bibfnamefont{V.}~\bibnamefont{Keppens}},
  \bibinfo{author}{\bibfnamefont{M.~A.} \bibnamefont{McGuire}},
  \bibinfo{author}{\bibfnamefont{D.}~\bibnamefont{Mandrus}},
  \bibinfo{author}{\bibfnamefont{B.~C.} \bibnamefont{Sales}}, \bibnamefont{and}
  \bibinfo{author}{\bibfnamefont{P.}~\bibnamefont{Thalmeier}},
  \bibinfo{journal}{Phys. Rev. Lett.} \textbf{\bibinfo{volume}{92}},
  \bibinfo{pages}{185502} (\bibinfo{year}{2004}).

\bibitem[{\citenamefont{Madsen and Santi}(2005)}]{GeorgPRB2005}
\bibinfo{author}{\bibfnamefont{G.~K.~H.} \bibnamefont{Madsen}}
  \bibnamefont{and} \bibinfo{author}{\bibfnamefont{G.}~\bibnamefont{Santi}},
  \bibinfo{journal}{Phys. Rev. B} \textbf{\bibinfo{volume}{72}},
  \bibinfo{pages}{220301} (\bibinfo{year}{2005}).

\bibitem[{\citenamefont{Biswas and Myles}(2007)}]{BiswasPRB2007}
\bibinfo{author}{\bibfnamefont{K.}~\bibnamefont{Biswas}} \bibnamefont{and}
  \bibinfo{author}{\bibfnamefont{C.~W.} \bibnamefont{Myles}},
  \bibinfo{journal}{Phys. Rev. B} \textbf{\bibinfo{volume}{75}},
  \bibinfo{pages}{245205} (\bibinfo{year}{2007}).

\bibitem[{\citenamefont{Avila et~al.}(2008)\citenamefont{Avila, Suekuni, Umeo,
  Fukuoka, Yamanaka, and Takabatake}}]{AvilaAPL2008}
\bibinfo{author}{\bibfnamefont{M.~A.} \bibnamefont{Avila}},
  \bibinfo{author}{\bibfnamefont{K.}~\bibnamefont{Suekuni}},
  \bibinfo{author}{\bibfnamefont{K.}~\bibnamefont{Umeo}},
  \bibinfo{author}{\bibfnamefont{H.}~\bibnamefont{Fukuoka}},
  \bibinfo{author}{\bibfnamefont{S.}~\bibnamefont{Yamanaka}}, \bibnamefont{and}
  \bibinfo{author}{\bibfnamefont{T.}~\bibnamefont{Takabatake}},
  \bibinfo{journal}{Appl. Phys. Lett.} \textbf{\bibinfo{volume}{92}},
  \bibinfo{pages}{041901} (\bibinfo{year}{2008}).

\bibitem[{\citenamefont{Suekuni et~al.}(2010)\citenamefont{Suekuni, Takasu,
  Hasegawa, Ogita, Udagawa, Avila, and Takabatake}}]{SuekuniPRB2010}
\bibinfo{author}{\bibfnamefont{K.}~\bibnamefont{Suekuni}},
  \bibinfo{author}{\bibfnamefont{Y.}~\bibnamefont{Takasu}},
  \bibinfo{author}{\bibfnamefont{T.}~\bibnamefont{Hasegawa}},
  \bibinfo{author}{\bibfnamefont{N.}~\bibnamefont{Ogita}},
  \bibinfo{author}{\bibfnamefont{M.}~\bibnamefont{Udagawa}},
  \bibinfo{author}{\bibfnamefont{M.~A.} \bibnamefont{Avila}}, \bibnamefont{and}
  \bibinfo{author}{\bibfnamefont{T.}~\bibnamefont{Takabatake}},
  \bibinfo{journal}{Phys. Rev. B} \textbf{\bibinfo{volume}{81}},
  \bibinfo{pages}{205207} (\bibinfo{year}{2010}).

\bibitem[{\citenamefont{Bentien et~al.}(2005)\citenamefont{Bentien, Nishibori,
  Paschen, and Iversen}}]{BentienPRB2005}
\bibinfo{author}{\bibfnamefont{A.}~\bibnamefont{Bentien}},
  \bibinfo{author}{\bibfnamefont{E.}~\bibnamefont{Nishibori}},
  \bibinfo{author}{\bibfnamefont{S.}~\bibnamefont{Paschen}}, \bibnamefont{and}
  \bibinfo{author}{\bibfnamefont{B.~B.} \bibnamefont{Iversen}},
  \bibinfo{journal}{Phys. Rev. B} \textbf{\bibinfo{volume}{71}},
  \bibinfo{pages}{144107} (\bibinfo{year}{2005}).

\bibitem[{\citenamefont{Mahan and Sofo}(1993)}]{MahanPRB1993}
\bibinfo{author}{\bibfnamefont{G.~D.} \bibnamefont{Mahan}} \bibnamefont{and}
  \bibinfo{author}{\bibfnamefont{J.~O.} \bibnamefont{Sofo}},
  \bibinfo{journal}{Phys. Rev. B} \textbf{\bibinfo{volume}{47}},
  \bibinfo{pages}{8050} (\bibinfo{year}{1993}).

\bibitem[{\citenamefont{Cooper}(1974)}]{CooperPRB1974}
\bibinfo{author}{\bibfnamefont{J.~R.} \bibnamefont{Cooper}},
  \bibinfo{journal}{Phys. Rev. B} \textbf{\bibinfo{volume}{9}},
  \bibinfo{pages}{2778} (\bibinfo{year}{1974}).

\bibitem[{\citenamefont{Mandrus et~al.}(2001)\citenamefont{Mandrus, Sales, and
  Jin}}]{MandrusPRB2001}
\bibinfo{author}{\bibfnamefont{D.}~\bibnamefont{Mandrus}},
  \bibinfo{author}{\bibfnamefont{B.~C.} \bibnamefont{Sales}}, \bibnamefont{and}
  \bibinfo{author}{\bibfnamefont{R.}~\bibnamefont{Jin}},
  \bibinfo{journal}{Phys. Rev. B} \textbf{\bibinfo{volume}{64}},
  \bibinfo{pages}{012302} (\bibinfo{year}{2001}).

\bibitem[{\citenamefont{Hermann et~al.}(2003)\citenamefont{Hermann, Jin,
  Schweika, Grandjean, Mandrus, Sales, and Long}}]{HermannPRL2003}
\bibinfo{author}{\bibfnamefont{R.~P.} \bibnamefont{Hermann}},
  \bibinfo{author}{\bibfnamefont{R.}~\bibnamefont{Jin}},
  \bibinfo{author}{\bibfnamefont{W.}~\bibnamefont{Schweika}},
  \bibinfo{author}{\bibfnamefont{F.}~\bibnamefont{Grandjean}},
  \bibinfo{author}{\bibfnamefont{D.}~\bibnamefont{Mandrus}},
  \bibinfo{author}{\bibfnamefont{B.~C.} \bibnamefont{Sales}}, \bibnamefont{and}
  \bibinfo{author}{\bibfnamefont{G.~J.} \bibnamefont{Long}},
  \bibinfo{journal}{Phys. Rev. Lett.} \textbf{\bibinfo{volume}{90}},
  \bibinfo{pages}{135505} (\bibinfo{year}{2003}).

\bibitem[{\citenamefont{Qiu et~al.}(2004)\citenamefont{Qiu, Swainson, Nolas,
  and White}}]{QiuPRB2004}
\bibinfo{author}{\bibfnamefont{L.}~\bibnamefont{Qiu}},
  \bibinfo{author}{\bibfnamefont{I.~P.} \bibnamefont{Swainson}},
  \bibinfo{author}{\bibfnamefont{G.~S.} \bibnamefont{Nolas}}, \bibnamefont{and}
  \bibinfo{author}{\bibfnamefont{M.~A.} \bibnamefont{White}},
  \bibinfo{journal}{Phys. Rev. B} \textbf{\bibinfo{volume}{70}},
  \bibinfo{pages}{035208} (\bibinfo{year}{2004}).

\bibitem[{\citenamefont{Umeo et~al.}(2005)\citenamefont{Umeo, Avila, Sakata,
  Suekuni, and Takabatake}}]{UmeoJPSJ2005}
\bibinfo{author}{\bibfnamefont{K.}~\bibnamefont{Umeo}},
  \bibinfo{author}{\bibfnamefont{M.~A.} \bibnamefont{Avila}},
  \bibinfo{author}{\bibfnamefont{T.}~\bibnamefont{Sakata}},
  \bibinfo{author}{\bibfnamefont{K.}~\bibnamefont{Suekuni}}, \bibnamefont{and}
  \bibinfo{author}{\bibfnamefont{T.}~\bibnamefont{Takabatake}},
  \bibinfo{journal}{J. Phys. Soc. Jpn.} \textbf{\bibinfo{volume}{74}},
  \bibinfo{pages}{2145} (\bibinfo{year}{2005}).

\bibitem[{\citenamefont{Suekuni et~al.}(2008)\citenamefont{Suekuni, Avila,
  Umeo, Fukuoka, Yamanaka, Nakagawa, and Takabatake}}]{SuekuniPRB2008}
\bibinfo{author}{\bibfnamefont{K.}~\bibnamefont{Suekuni}},
  \bibinfo{author}{\bibfnamefont{M.~A.} \bibnamefont{Avila}},
  \bibinfo{author}{\bibfnamefont{K.}~\bibnamefont{Umeo}},
  \bibinfo{author}{\bibfnamefont{H.}~\bibnamefont{Fukuoka}},
  \bibinfo{author}{\bibfnamefont{S.}~\bibnamefont{Yamanaka}},
  \bibinfo{author}{\bibfnamefont{T.}~\bibnamefont{Nakagawa}}, \bibnamefont{and}
  \bibinfo{author}{\bibfnamefont{T.}~\bibnamefont{Takabatake}},
  \bibinfo{journal}{Phys. Rev. B} \textbf{\bibinfo{volume}{77}},
  \bibinfo{pages}{235119} (\bibinfo{year}{2008}).

\bibitem[{\citenamefont{Lortz et~al.}(2008)\citenamefont{Lortz, Viennois,
  Petrovic, Wang, Toulemonde, Meingast, Koza, Mutka, Bossak, and
  Miguel}}]{LortzPRB2008}
\bibinfo{author}{\bibfnamefont{R.}~\bibnamefont{Lortz}},
  \bibinfo{author}{\bibfnamefont{R.}~\bibnamefont{Viennois}},
  \bibinfo{author}{\bibfnamefont{A.}~\bibnamefont{Petrovic}},
  \bibinfo{author}{\bibfnamefont{Y.}~\bibnamefont{Wang}},
  \bibinfo{author}{\bibfnamefont{P.}~\bibnamefont{Toulemonde}},
  \bibinfo{author}{\bibfnamefont{C.}~\bibnamefont{Meingast}},
  \bibinfo{author}{\bibfnamefont{M.~M.} \bibnamefont{Koza}},
  \bibinfo{author}{\bibfnamefont{H.}~\bibnamefont{Mutka}},
  \bibinfo{author}{\bibfnamefont{A.}~\bibnamefont{Bossak}}, \bibnamefont{and}
  \bibinfo{author}{\bibfnamefont{A.~S.} \bibnamefont{Miguel}},
  \bibinfo{journal}{Phys. Rev. B} \textbf{\bibinfo{volume}{77}},
  \bibinfo{pages}{224507} (\bibinfo{year}{2008}).

\bibitem[{\citenamefont{Dahm and Ueda}(2007)}]{DahmPRL2007}
\bibinfo{author}{\bibfnamefont{T.}~\bibnamefont{Dahm}} \bibnamefont{and}
  \bibinfo{author}{\bibfnamefont{K.}~\bibnamefont{Ueda}},
  \bibinfo{journal}{Phys. Rev. Lett.} \textbf{\bibinfo{volume}{99}},
  \bibinfo{pages}{187003} (\bibinfo{year}{2007}).

\bibitem[{\citenamefont{Y.~Rodriguez et~al.}(2010)\citenamefont{Y.~Rodriguez,
  Zheng, Saribaev, and Ross}}]{SergioMRS2010}
\bibinfo{author}{\bibfnamefont{S.}~\bibnamefont{Y.~Rodriguez}},
  \bibinfo{author}{\bibfnamefont{X.}~\bibnamefont{Zheng}},
  \bibinfo{author}{\bibfnamefont{L.}~\bibnamefont{Saribaev}}, \bibnamefont{and}
  \bibinfo{author}{\bibfnamefont{J.~H.} \bibnamefont{Ross},
  \bibfnamefont{Jr.}}, \bibinfo{journal}{Proc. Mater. Res. Soc.}
  \textbf{\bibinfo{volume}{1267}}, \bibinfo{pages}{DD04}
  (\bibinfo{year}{2010}).

\bibitem[{\citenamefont{Ziman}(1972)}]{Bloch-Gruneisen-Ziman}
\bibinfo{author}{\bibfnamefont{J.~M.} \bibnamefont{Ziman}},
  \emph{\bibinfo{title}{Principles of the Theory of Solids}}
  (\bibinfo{publisher}{Cambridge University Press},
  \bibinfo{address}{Cambridge}, \bibinfo{year}{1972}).

\bibitem[{\citenamefont{Bulusu and Walker}(2008)}]{BulusuSM2008}
\bibinfo{author}{\bibfnamefont{A.}~\bibnamefont{Bulusu}} \bibnamefont{and}
  \bibinfo{author}{\bibfnamefont{D.~G.} \bibnamefont{Walker}},
  \bibinfo{journal}{Superlattices and Microst.} \textbf{\bibinfo{volume}{44}},
  \bibinfo{pages}{1 } (\bibinfo{year}{2008}).

\bibitem[{\citenamefont{Kittel}(1986)}]{Kittel}
\bibinfo{author}{\bibfnamefont{C.}~\bibnamefont{Kittel}},
  \emph{\bibinfo{title}{Introduction to Solid State Physics, 6th Edition}}
  (\bibinfo{publisher}{John Wiley \& Sons, Inc.}, \bibinfo{address}{New York},
  \bibinfo{year}{1986}).

\bibitem[{\citenamefont{Grimvall}(1999)}]{Grimvall}
\bibinfo{author}{\bibfnamefont{G.}~\bibnamefont{Grimvall}},
  \emph{\bibinfo{title}{Thermophysical Properties of Materials}}
  (\bibinfo{publisher}{North-Holland}, \bibinfo{address}{Amsterdam},
  \bibinfo{year}{1999}).

\bibitem[{\citenamefont{Dong et~al.}(2000)\citenamefont{Dong, Sankey,
  Ramachandran, and McMillan}}]{DongJAP2000}
\bibinfo{author}{\bibfnamefont{J.}~\bibnamefont{Dong}},
  \bibinfo{author}{\bibfnamefont{O.~F.} \bibnamefont{Sankey}},
  \bibinfo{author}{\bibfnamefont{G.~K.} \bibnamefont{Ramachandran}},
  \bibnamefont{and} \bibinfo{author}{\bibfnamefont{P.~F.}
  \bibnamefont{McMillan}}, \bibinfo{journal}{J. Appl. Phys.}
  \textbf{\bibinfo{volume}{87}}, \bibinfo{pages}{7726} (\bibinfo{year}{2000}).

\bibitem[{\citenamefont{Reny et~al.}(2002)\citenamefont{Reny, San-Miguel,
  Guyot, Masenelli, M\'elinon, Saviot, Yamanaka, Champagnon, Cros, Pouchard
  et~al.}}]{RenyPRB2002}
\bibinfo{author}{\bibfnamefont{E.}~\bibnamefont{Reny}},
  \bibinfo{author}{\bibfnamefont{A.}~\bibnamefont{San-Miguel}},
  \bibinfo{author}{\bibfnamefont{Y.}~\bibnamefont{Guyot}},
  \bibinfo{author}{\bibfnamefont{B.}~\bibnamefont{Masenelli}},
  \bibinfo{author}{\bibfnamefont{P.}~\bibnamefont{M\'elinon}},
  \bibinfo{author}{\bibfnamefont{L.}~\bibnamefont{Saviot}},
  \bibinfo{author}{\bibfnamefont{S.}~\bibnamefont{Yamanaka}},
  \bibinfo{author}{\bibfnamefont{B.}~\bibnamefont{Champagnon}},
  \bibinfo{author}{\bibfnamefont{C.}~\bibnamefont{Cros}},
  \bibinfo{author}{\bibfnamefont{M.}~\bibnamefont{Pouchard}},
  \bibnamefont{et~al.}, \bibinfo{journal}{Phys. Rev. B}
  \textbf{\bibinfo{volume}{66}}, \bibinfo{pages}{014532}
  (\bibinfo{year}{2002}).

\bibitem[{\citenamefont{Hermann et~al.}(2005)\citenamefont{Hermann, Schweika,
  Leupold, R\"uffer, Nolas, Grandjean, and Long}}]{HermannPRB2005}
\bibinfo{author}{\bibfnamefont{R.~P.} \bibnamefont{Hermann}},
  \bibinfo{author}{\bibfnamefont{W.}~\bibnamefont{Schweika}},
  \bibinfo{author}{\bibfnamefont{O.}~\bibnamefont{Leupold}},
  \bibinfo{author}{\bibfnamefont{R.}~\bibnamefont{R\"uffer}},
  \bibinfo{author}{\bibfnamefont{G.~S.} \bibnamefont{Nolas}},
  \bibinfo{author}{\bibfnamefont{F.}~\bibnamefont{Grandjean}},
  \bibnamefont{and} \bibinfo{author}{\bibfnamefont{G.~J.} \bibnamefont{Long}},
  \bibinfo{journal}{Phys. Rev. B} \textbf{\bibinfo{volume}{72}},
  \bibinfo{pages}{174301} (\bibinfo{year}{2005}).

\bibitem[{\citenamefont{Johnsen et~al.}(2010)\citenamefont{Johnsen,
  Christensen, Thomsen, Madsen, and Iversen}}]{JohnsonPRB2010}
\bibinfo{author}{\bibfnamefont{S.}~\bibnamefont{Johnsen}},
  \bibinfo{author}{\bibfnamefont{M.}~\bibnamefont{Christensen}},
  \bibinfo{author}{\bibfnamefont{B.}~\bibnamefont{Thomsen}},
  \bibinfo{author}{\bibfnamefont{G.~K.~H.} \bibnamefont{Madsen}},
  \bibnamefont{and} \bibinfo{author}{\bibfnamefont{B.~B.}
  \bibnamefont{Iversen}}, \bibinfo{journal}{Phys. Rev. B}
  \textbf{\bibinfo{volume}{82}}, \bibinfo{pages}{184303}
  (\bibinfo{year}{2010}).

\bibitem[{\citenamefont{Christensen et~al.}(2008)\citenamefont{Christensen,
  Abrahamsen, Christensen, Juranyi, Andersen, Lefmann, Andreasson, Bahl, and
  Iversen}}]{ChistensenNatMater2008}
\bibinfo{author}{\bibfnamefont{M.}~\bibnamefont{Christensen}},
  \bibinfo{author}{\bibfnamefont{A.~B.} \bibnamefont{Abrahamsen}},
  \bibinfo{author}{\bibfnamefont{N.~B.} \bibnamefont{Christensen}},
  \bibinfo{author}{\bibfnamefont{F.}~\bibnamefont{Juranyi}},
  \bibinfo{author}{\bibfnamefont{N.~H.} \bibnamefont{Andersen}},
  \bibinfo{author}{\bibfnamefont{K.}~\bibnamefont{Lefmann}},
  \bibinfo{author}{\bibfnamefont{J.}~\bibnamefont{Andreasson}},
  \bibinfo{author}{\bibfnamefont{C.~R.~H.} \bibnamefont{Bahl}},
  \bibnamefont{and} \bibinfo{author}{\bibfnamefont{B.~B.}
  \bibnamefont{Iversen}}, \bibinfo{journal}{Nat. Mater.}
  \textbf{\bibinfo{volume}{7}}, \bibinfo{pages}{811} (\bibinfo{year}{2008}).

\bibitem[{\citenamefont{Christensen et~al.}(2006)\citenamefont{Christensen,
  Lock, Overgaard, and Iversen}}]{ChristensenJACS2006}
\bibinfo{author}{\bibfnamefont{M.}~\bibnamefont{Christensen}},
  \bibinfo{author}{\bibfnamefont{N.}~\bibnamefont{Lock}},
  \bibinfo{author}{\bibfnamefont{J.}~\bibnamefont{Overgaard}},
  \bibnamefont{and} \bibinfo{author}{\bibfnamefont{B.~B.}
  \bibnamefont{Iversen}}, \bibinfo{journal}{J. Am. Chem. Soc.}
  \textbf{\bibinfo{volume}{128}}, \bibinfo{pages}{15657}
  (\bibinfo{year}{2006}).

\bibitem[{\citenamefont{Roudebush et~al.}(2012)\citenamefont{Roudebush, de~la
  Cruz, Chakoumakos, and Kauzlarich}}]{RoudebushInorgChem2012}
\bibinfo{author}{\bibfnamefont{J.~H.} \bibnamefont{Roudebush}},
  \bibinfo{author}{\bibfnamefont{C.}~\bibnamefont{de~la Cruz}},
  \bibinfo{author}{\bibfnamefont{B.~C.} \bibnamefont{Chakoumakos}},
  \bibnamefont{and} \bibinfo{author}{\bibfnamefont{S.~M.}
  \bibnamefont{Kauzlarich}}, \bibinfo{journal}{Inorg. Chem.}
  \textbf{\bibinfo{volume}{51}}, \bibinfo{pages}{1805} (\bibinfo{year}{2012}).

\bibitem[{\citenamefont{Lee et~al.}(2008)\citenamefont{Lee, Yoshizawa, Avila,
  Hase, Kihou, and Takabatake}}]{LeeJPSJ2008}
\bibinfo{author}{\bibfnamefont{C.-H.} \bibnamefont{Lee}},
  \bibinfo{author}{\bibfnamefont{H.}~\bibnamefont{Yoshizawa}},
  \bibinfo{author}{\bibfnamefont{M.~A.} \bibnamefont{Avila}},
  \bibinfo{author}{\bibfnamefont{I.}~\bibnamefont{Hase}},
  \bibinfo{author}{\bibfnamefont{K.}~\bibnamefont{Kihou}}, \bibnamefont{and}
  \bibinfo{author}{\bibfnamefont{T.}~\bibnamefont{Takabatake}},
  \bibinfo{journal}{J. Phys. Soc. Jpn.} \textbf{\bibinfo{volume}{77SA}},
  \bibinfo{pages}{260} (\bibinfo{year}{2008}).

\bibitem[{\citenamefont{Ramos and Buchenau}(1998)}]{RamosSPM}
\bibinfo{author}{\bibfnamefont{M.~A.} \bibnamefont{Ramos}} \bibnamefont{and}
  \bibinfo{author}{\bibfnamefont{U.}~\bibnamefont{Buchenau}},
  \emph{\bibinfo{title}{Tunneling Systems in Amorphous and Crystalline Solids}}
  (\bibinfo{publisher}{Springer}, \bibinfo{address}{Berlin},
  \bibinfo{year}{1998}), chap.~\bibinfo{chapter}{9}.

\bibitem[{\citenamefont{Lindqvist et~al.}(1997)\citenamefont{Lindqvist,
  Yamamuro, Tsukushi, and Matsuo}}]{LindqvistJCP1997}
\bibinfo{author}{\bibfnamefont{A.}~\bibnamefont{Lindqvist}},
  \bibinfo{author}{\bibfnamefont{O.}~\bibnamefont{Yamamuro}},
  \bibinfo{author}{\bibfnamefont{I.}~\bibnamefont{Tsukushi}}, \bibnamefont{and}
  \bibinfo{author}{\bibfnamefont{T.}~\bibnamefont{Matsuo}},
  \bibinfo{journal}{J. Chem. Phys.} \textbf{\bibinfo{volume}{107}},
  \bibinfo{pages}{5103} (\bibinfo{year}{1997}).

\end{thebibliography}
\end{document}